# Possible Relevance of Odd Frequency Pairing to Heavy Fermion Superconductivity

P. Coleman and E. Miranda

*Serin Physics Laboratory, Rutgers University, PO Box 849, Piscataway NJ, 08855.*

What is the character of the gapless quasiparticles in heavy fermion superconductors (HFSC)? We discuss an odd-frequency pairing interpretation of HFSC which leads to a two component model for the quasiparticle excitations. In this picture, line zeroes of unpaired electrons may coexist with gapless surfaces of paired electrons, with vanishing spin and charge coherence factors.

After more than a decade of debate over the nature of heavy fermion superconductivity (HFSC), this extraordinary phenomenon still defies a consistent theoretical description. These superconductors are in essence, "spin superconductors", where thermodynamics implicates the active participation of local moment degrees of freedom in the condensation process.

Experimentally, there is ample evidence for the presence of gapless quasiparticles in the superconducting state[1]. In $UPt_3$, for example, $\mu$SR and transverse ultrasound absorption measurements indicate the presence of lines of gap zeroes in the basal plane[2,3]. Phenomenological theories of HFSC have modeled the quasiparticle (QP) properties of the HFSC in terms of a BCS superconductor with lines of node zeroes in the gap function.[4] These types of scenarios are unable to reconcile

⋄ The almost isotropic residual thermal conductivity observed in single crystals of $UPt_3$ with the assumption that QP excitations are concentrated in the basal plane.[5]

⋄ The absence of a large linear Korringa NMR relaxation rate ($1/T_1 \propto T$) in the superconducting state, despite sizable linear specific heat capacities in the superconducting phase[6,7].

Here we discuss an alternative interpretation of these observations, based on the idea that the HFSC gap function is an odd function of the frequency[8–10]

$$\Delta_{\vec{k}}(\omega) = -\Delta_{\vec{k}}(-\omega)$$

In an anisotropic BCS superconductor, gapless quasiparticles are unpaired, for the Bogoliubov coefficients in the QP operators

$$a^\dagger_{\vec{k}\uparrow} = u_{\vec{k}} c^\dagger_{\vec{k}\uparrow} + v_{\vec{k}} c_{-\vec{k}\downarrow}$$

are zero, or unity when the QP energy $E_{\vec{k}} = 0$. For an odd frequency superconductor, $|u_{\vec{k}}|^2 = |v_{\vec{k}}|^2 = \frac{1}{2}$ when $E_{\vec{k}} = 0$, so the gapless quasiparticles are *paired*; the equal weight of particle and hole implies that their charge coherence factors identically vanish,

$$\langle \vec{k} | \hat{Q} | \vec{k} \rangle \propto u_{\vec{k}}^2 - v_{\vec{k}}^2 = 0.$$

Such "neutral Fermi surfaces" of paired quasiparticles can coexist with more conventional point and line nodes of unpaired electrons leading to a *"two component"* description of the QP fluid.

Odd frequency superconductivity cannot proceed via the conventional two-body Cooper instability. We propose that HFSC is driven by *the development and condensation of three-body bound states between conduction electrons and localized f-spins*. Unlike pair condensation, three-body condensation must leave behind a neutral, spinless fermion associated with the residual fermionic fluctuations of the bound-state. This Ansatz is implemented using a three-body contraction to develop a mean field theory

$$\overline{(\vec{\sigma} \cdot \vec{S}_j) \psi_{\Gamma j}} = \mathcal{Z}_j \hat{\phi}_j \quad (1)$$

where $j$ is a site index, $\vec{S}_j$ describes spin fluctuations within a low-lying crystal field doublet $\Gamma$, $\psi_{\Gamma j}$ denotes a conduction electron in a Wannier state with symmetry $\Gamma$ and $\mathcal{Z}_j$ is a two component spinor describing the three body condensate. The *operator* $\hat{\phi}_j = \hat{\phi}_j^\dagger$ is a real, or Majorana fermion that represents the appearance of a bound-state pole in the three body channel.

We illustrate this idea within a Kondo lattice model

$$H = \sum_{\vec{k}} \epsilon_{\vec{k}} \psi^\dagger_{\vec{k}} \psi_{\vec{k}} + \sum_j H_{int}(j) \quad (2)$$

where $\psi_{\vec{k}}$ is a conduction electron spinor. We write the Kondo exchange interaction in the form

$$H_{int}(j) = -J \xi_j^\dagger \xi_j \quad (3)$$

where

$$\xi_j = (\vec{\sigma} \cdot \vec{S}_j) \psi_{\Gamma j} \quad (4)$$

is a three-body spinor composed of an f-spin and conduction electron $\psi_{\Gamma j} = \sum_{\vec{k}} \gamma_\Gamma(\vec{k}) \psi_{\vec{k}} e^{i \vec{k} \cdot \vec{R}_j}$ in a Wannier state of symmetry $\Gamma$.

The three-body Ansatz is then written

$$-J \hat{\xi}_j = 2 V_j \hat{\phi}_j + \text{fluctuations} \quad (5)$$

After inserting this into (3)

$$H_{int}(j) \rightarrow H_{mft}(j) = 2(V_j^\dagger \phi_j \xi_j + \text{H.c}) + 2V_j^\dagger V_j/J, \quad (6)$$

a marvelous simplification occurs: the fusion of the bound-state fermion $\hat{\phi}_j$ and local moment $\vec{S}_j$ forms a triplet of real "spin fermions" $\vec{\eta}(j) = 2\phi_j \vec{S}_j$, ($\{\eta^a(i), \eta^b(j)\} = \delta^{ab}\delta_{ij}$). $H_{int}(j)$ is thus a fermion bilinear

$$H_{mft}(j) = (V_j^\dagger (\vec{\sigma} \cdot \vec{\eta}_j)\psi_{\Gamma j} + \text{H.c}) + 2V_j^\dagger V_j/J \quad (7)$$

Consider a toy model on a cubic lattice with. Let $\gamma_\Gamma(\vec{k}) = \sin k_z$[11] vanish in the basal plane. For the saddle-point solution $V_j^\dagger = e^{i\vec{Q}\cdot\vec{R}_j/2}V_o(1,0)$, $\vec{Q} = (\pi, \pi, \pi)$, the conduction electron Green function has the form

$$\mathcal{G}(\vec{k}, \omega) = [\omega - \tilde{\epsilon}_{\vec{k}} + \mu\tau_3 - \Delta_{\vec{k}}(\omega)\mathcal{P}]^{-1} \quad (8)$$

where $\mathcal{P} = \frac{1}{4}(3 - \vec{\sigma}\cdot\vec{\tau})$ is a projection operator, $\vec{\tau}$ denotes isospin operators, $\tilde{\epsilon}_{\vec{k}} = \frac{1}{2}(\epsilon_{\vec{k}-\vec{Q}/2} - \epsilon_{\vec{k}+\vec{Q}/2})$. The gap function

$$\Delta_{\vec{k}}(\omega) = \frac{V_o^2}{\omega}[\tilde{\gamma}_\Gamma(\vec{k})]^2,$$

where $\tilde{\gamma}_\Gamma(\vec{k}) = \gamma_\Gamma(\vec{k}+\vec{Q}/2)$, acquires the nodal symmetry of the crystal field state $\Gamma$.[12] This simple model[11] gives rise to two sets of gapless quasiparticles (Fig. 1.):

- A neutral Fermi surface $\tilde{\epsilon}_{\vec{k}} = 0$ of gapless paired quasiparticles with vanishing spin and charge coherence factors.

- A line of unpaired gapless quasiparticles in the basal plane $k_z = 0$, corresponding to the node in the crystal field wave function $\tilde{\gamma}_\Gamma(\vec{k})$. These excitations have a linear density of states.

Local probes of the quasiparticle fluid are unable to distinguish between these two components, both of which would give rise to $T^3$ NMR relaxation rate. Momentum space probes, such as the anisotropic penetration depth or the transverse ultrasound absorption are sensitive to the current fluctuations. For example, the temperature dependence of the superfluid stiffness tensor

$$\Delta \rho_{ab}(T) \propto -\int d\omega \frac{\partial f(\omega)}{\partial \omega} \sigma_{ab}(\omega)$$

where

$$\sigma_{ab}(\omega) = \sum_{\vec{k}\lambda} \langle \vec{k}\lambda|j_a|\vec{k}\lambda\rangle \langle \vec{k}\lambda|j_a|\vec{k}\lambda\rangle \delta(E_{\vec{k}\lambda} - \omega) \quad (9)$$

measures the spectrum of current fluctuations. In the basal plane, the current matrix elements $\langle \vec{k}\lambda|j_a|\vec{k}\lambda\rangle$ are unity, and this function reflects the linear density of states of unpaired quasiparticles $\sigma_{xx}(\omega) \propto \omega$. Along the z axis, these matrix elements grow linearly with energy, but

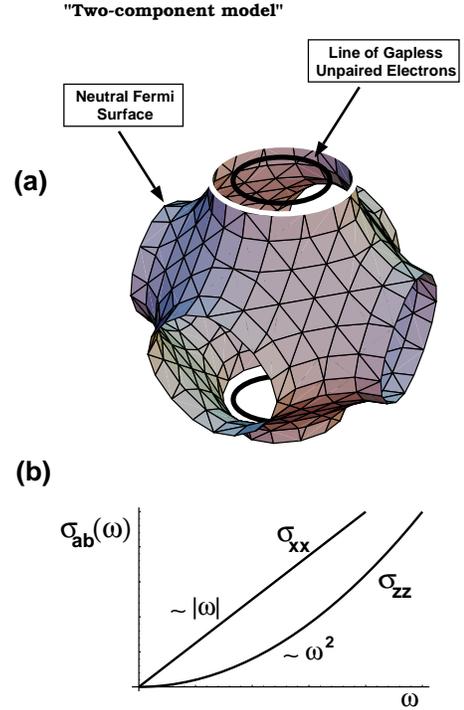

Fig. 1. (a) Neutral Fermi surface and line of gapless unpaired electrons obtained in simple model; (b) schematic illustration of the anisotropic conductivity predicted in this model in ($\sigma_{xx}$), and out of ($\sigma_{zz}$) the basal plane.

the density of states is constant, giving rise to a quadratic frequency dependence of $\sigma_{zz}(\omega) \propto \omega^2$, as illustrated in Fig. 1.

To conclude, we have outlined how a three-fermion condensation into states with definite crystal symmetry generates an anisotropic odd-frequency superconductor. Nodes in the atomic crystal field wave function give rise to corresponding nodes in the gap function, leading to a coexistence of line nodes of electrons with gapless surfaces of paired, neutral quasiparticles. Confirmation of the absence of Korringa relaxation and anisotropy in the residual thermal conductivity in samples with large linear specific heat would provide strong support in favor of an odd-frequency pairing scenario.

We would like to thank L. Ioffe and A. Tsvelik for discussions related to this work. We would like to thank the International Center for Theoretical Physics, Trieste, and the Aspen Center for Physics, where part of this work was carried out. This work was supported by NSF grant DMR-93-12138 and CNPq, Brazil.

---